\documentclass{JHEP3}

\usepackage{graphics}
\usepackage{rotating}
\usepackage{amssymb}
\usepackage{pstricks}

\makeatletter
\@addtoreset{equation}{section}
\makeatother

\def\dalemb#1#2{{\vbox{\hrule height .#2pt
        \hbox{\vrule width.#2pt height#1pt \kern#1pt
                \vrule width.#2pt}
        \hrule height.#2pt}}}

\def\tA{\widetilde A}

\def\CC{{\mathbb C}}
\def\RR{{\mathbb R}}

\def\0{{\sst{(0)}}}
\def\1{{\sst{(1)}}}
\def\2{{\sst{(2)}}}
\def\3{{\sst{(3)}}}
\def\4{{\sst{(4)}}}
\def\5{{\sst{(5)}}}
\def\6{{\sst{(6)}}}
\def\7{{\sst{(7)}}}
\def\8{{\sst{(8)}}}

\def\tA{\widetilde A}

\def\Z{\rlap{\sf Z}\mkern3mu{\sf Z}}
\def\R{\rlap{\rm I}\mkern3mu{\rm R}}
\def\G{{\cal G}}
\def\A{{\cal A}}

\def\S{{\cal S}}

\def\T{{\cal T}}

\def\N{{\cal N}}

\def\P{{\cal P}}

 \def\bd{\begin{document}} \def\ed{\end{document}}
\def\ds{\documentstyle} \let\fr=\frac \let\bl=\bigl \let\br=\bigr
\let\Br=\Bigr \let\Bl=\Bigl
\let\bm=\bibitem
\let\na=\nabla
\let\pa=\partial \let\ov=\overline
\newcommand{\be}{\begin{equation}}
\newcommand{\ee}{\end{equation}}
\def\ba{\begin{array}}
\def\ea{\end{array}}
\def\ft#1#2{{\textstyle{{\scriptstyle #1}\over {\scriptstyle #2}}}}
\def\fft#1#2{{#1 \over #2}}
\def\del{\partial}
\def\sst#1{{\scriptscriptstyle #1}}
\def\oneone{\rlap 1\mkern4mu{\rm l}}
\def\ie{{\it i.e.\ }}
\def\via{{\it via}}
\def\semi{{\ltimes}}
\def\v{{\cal V}}
\def\str{{\rm str}}
\def\Dm{{{D_{\sst{max}}}}}

\newcommand{\ho}[1]{$\, ^{#1}$}
\newcommand{\hoch}[1]{$\, ^{#1}$}
\newcommand{\bea}{\begin{eqnarray}}
\newcommand{\eea}{\end{eqnarray}}
\newcommand{\ra}{\rightarrow}
\newcommand{\lra}{\longrightarrow}
\newcommand{\Lra}{\Leftrightarrow}
\newcommand{\bp}{\tilde \beta^\prime}
\newcommand{\tr}{{\rm tr} }
\newcommand{\Tr}{{\rm Tr} }

\newcommand{\CP}{{\mathbb C}{\mathbb P}}

\newcommand{\RP}{{\mathbb R}{\mathbb P}}

\renewcommand{\floatpagefraction}{0.05}

\newcommand{\NP}{Nucl. Phys. }
\newcommand{\ens}{\it Laboratoire de Physique Th\'eorique de l'\'Ecole
Normale Sup\'erieure\\
24 Rue Lhomond - 75231 Paris CEDEX 05}

\title{Real Borcherds Superalgebras and M-theory}

\author{Pierre Henry-Labord\`ere$^{\spadesuit, \heartsuit}$, Bernard Julia$^{\heartsuit}$ and Louis Paulot$^{\heartsuit}$   \\
${\tiny \spadesuit}$ Queen Mary and Westfield College, University of London \\
Mile End Road, London E1 4NS\\
${\tiny \heartsuit}$ Laboratoire de Physique Th{\'e}orique
de l'Ecole Normale Sup{\'e}rieure\thanks{ UMR 8549 du Centre National
de la Recherche Scientifique et de l'\'Ecole Normale Sup\'erieure}\\
24  rue Lhomond, 75231 Paris Cedex 05, France\\
\email{p.henry-labordere@qmul.ac.uk}\\
\email{bernard.julia@lpt.ens.fr}\\
\email{paulot@lpt.ens.fr}}

\abstract{
The correspondence between del Pezzo surfaces and field theory models,
discussed  in \cite{inv} and in
\cite{pbl} over the complex numbers or for split real forms, is
extended to other real forms, in particular to those
compatible with supersymmetry.
Specifically, all theories of the Magic triangle
\cite{jul1} that reduce to the pure supergravities in four
dimensions correspond to singular real del Pezzo surfaces and the
same is true for the Magic
square of ${\cal N}=2$ SUGRAS
\cite{gun}. 

A real del Pezzo surface is the invariant set under an
antilinear involution of a complex one.
This conjugation induces an involution of the Picard group that
preserves the anticanonical class and the  intersection form.
The known non-split U-duality algebras are embedded into
Borcherds superalgebras defined by their Cartan matrix (minus the intersection
form) and fixed by the anti-involution.
These data may be described by Tits-Satake bicoloured superdiagrams.

As in the split case, oxidation results from blowing down disjoint 
real ${\mathbb P}^1$'s of self-intersection $-1$.
The singular del Pezzo
surfaces of interest are obtained by degenerating
regular surfaces upon contraction of 
real curves of self-intersection $-2$.
We use the finite classification of real simple singularities 
to exhibit the relevant 
normal surfaces. We also give a general construction of more magic triangles 
like a type I split magic triangle and  prove their
(approximate) symmetry  with respect to their diagonal, this symmetry argument 
was announced in our previous paper for the split case.
}

\preprint{LPTENS-02/43\\hep-th/0212346}

\begin{document}

\addtocontents{toc}{\protect\setcounter{tocdepth}{2}}

\section{Introduction}

Finite dimensional Lie theory has been crucial for the construction of 
the standard model. 
In the same way, symmetry algebras will be crucial to
understand the structure of M-theory. Recently, we have embedded the
U-duality algebra $E_{n|n}$ for eleven-dimensional supergravity
compactified on a torus $T^n$ in a Borcherds superalgebra using a
connection with the del Pezzo surfaces \cite{pbl}.  This huge algebra in turn
could be truncated to exhibit the superalgebra of symmetries of \cite{cjlp2}.

The classification of projective algebraic curves (compact connected Riemann surfaces)
leads to the genus invariant and to the list: Riemann sphere, elliptic (plane) curves and
quotients of hyperbolic half-plane by cocompact discrete subgroups of $SL(2,\RR)$.
The classification of smooth 
projective algebraic surfaces has also been known for a while, but 
the Enriques-Kodaira theory is much richer than uniformisation theory. One distinguishes
\cite{bea,rei} them first by their Kodaira dimension $\kappa$ which can be $-\infty, 0 , 1 , 2$. The
first case is that of ruled surfaces (ie birationally equivalent to the product of a curve $\Sigma_g$ 
by a projective line). The birational class of $\CP^2$ is the
 set of rational surfaces ($g=0$). The 
other ruled surfaces are obtained by a set of blowups from projective line bundles on curves.
The latter are  called their ``minimal models'' for that reason. Among rational surfaces one 
meets first the del Pezzo surfaces but also the Veronese surface. The case 
of null Kodaira dimension encompasses K3 surfaces and their quotients the 
Enriques surfaces  as well as the abelian and the bielliptic surfaces. Non rational 
elliptic pencils have Kodaira dimension one, and the rest are the so-called general 
surfaces. For nonruled surfaces, the minimal models are unique and their classification 
is the same as the classification  up to birational equivalence.
For rational surfaces, one notes that $\CP^2$ and $\CP^1 \times \CP^1$ are both minimal, 
birationally equivalent to each other but nonisomorphic surfaces.  
 
The Picard group or second homology
of the del Pezzo surfaces defines a nice lattice connected with the root
lattice of the Borcherds superalgebras. These enlarged algebras require the
use of a new formulation of M-theory which introduces the fields and
their duals \cite{cjlp2}. The equations of motion for the various
$p$-forms should then be written as a self-duality equation based on
a supercoset $G/K_G$ with $G$ the supergroup corresponding to the
Borcherds superalgebra and $K_G$ its maximal "compact" subgroup, defined
as the fixed elements under the Cartan-Chevalley-Serre involution. The other
U-duality algebras appearing in the oxidation of $E_{n|n}$ cosets from
3 dimensions (for $n\le 7$) form another (``Split Magic'')  triangle,  SM$\Delta$, 
they have also been extended to
split Borcherds superalgebras and connected with (singular) normal del Pezzo
surfaces in \cite{pbl}.  The reflection symmetry along the diagonal of SM$\Delta$ 
was proved there.

Now the supergravity theories that give pure supergravity in four
dimension with less than $\N =8$ supersymmetries form the original Magic
Triangle M$\Delta$ \cite{jul1}, the U-duality algebras include non-split
real forms of simple Lie algebras.  We will show in this paper how
to extend these non-split forms to non-split Borcherds
superalgebras too. The conjugation defining a real form of the 
Borcherds superalgebra will be connected with the real structure of the
corresponding  real normal del Pezzo surface, defined by a complex
conjugation which induces an involution of the Picard group that
preserves the anticanonical class. These real forms of Borcherds
superalgebras will then be characterized by a Satake superdiagram.
In the first part ofi this paper, as a warm-up exercise, 
we give the complete generalizable  proof of
the symmetry of the Split Magic triangle and construct yet another
symmetric triangle starting from type I supergravity. 

Then we shall review the theory of real del Pezzo surfaces. Section 4 is on real forms of Lie 
algebras and section 5 on the (nonsplit) Magic triangle and Magic square, there we reproduce
supergravity models by adding by hand appropriate singularities on the surfaces. In the 
conclusion and appendix we emphasize some open questions, in particular the need to 
understand better supersymmetry breaking in this context as well as some peculiarity of $\N=2$ 
pure SUGRA.

\section{Symmetry of triangles}

It was noticed in \cite{cjlp} that the ``Split'' Magic Triangle (SM$\Delta$)
constructed by oxidizing some split non-linear $\sigma$-models has a
symmetry across the diagonal (Table 1). There are 5 bifurcations which we discovered in
\cite{pbl}, they appear in the text of that paper but unfortunately not on the table so we
do it here. The bifurcations reflect the lack of uniqueness of the minimal model.
For completeness we should recall that another 
magic triangle related to exceptional groups extending the ideas of Freudenthal and 
Tits has been proposed by P. Cvitanovic \cite{cvi}, however it does involve 
non-simply laced algebras and does not fit with our more primitive constructions yet. 
Actually it was noted long ago by the second author that the Magic square of 
\cite{gun} corresponds to Tits geometries of types $F_4$ resp $C_3$, $A_2$, $A_1$. So in 
effect 
non-simply laced algebras do enter in the theory of real forms of simply laced ones. 

 In \cite{pbl} we associated to each
of these theories a del Pezzo surface. These del Pezzo's are
possibly singular, but the singularity is at worst Du Val, which
means that there are isolated singular points which can be
resolved in a set of (-2)-curves intersecting according to an ADE
Cartan matrix (in fact A type singularities suffice). 
Before translating the  symmetry of the triangle into Lie
superalgebra language, we will first explain it at the level of
algebraic geometry of del Pezzo surfaces.

\begin{sidewaystable}[htbp]
\begin{center}
\begin{tabular}{|c|c|c|c|c|c|c|c|c|c|}\hline
&$n=8$&$n=7$&$n=6$&$n=5$&$n=4$&$n=3$&$n=2$&$n=1$&$n=0$\\ \hline
$d=11$ & + & & & & & & & &\\ \hline
$d=10$ &$\R {\it or} A_1$ & +  &&&&&&& \\ \hline
$d=9$ & $\R\times A_1$ & $\R$ & & &&&&& \\ \hline
$d=8$ & $A_1\times A_2$ & $\R \times A_1$ {\it or} $A_2$& $A_1$ & &&&&&\\ \hline
$d=7$ & $E_4$ & $\R \times A_2$ & $\R \times A_1$ & $\R$ & + & & &&\\ \hline
$d=6$ & $E_5$ & $ A_1\times A_3$ & $\R\times A_1^2$
        & $ \R ^2$ {\it or} $A_1^2$& $\R$ & & & &    \\ \hline
$d=5$ & $E_6$ & $A_5$ & $A_2^2$ & $
       \R\times A_1^2$ & $\R \times A_1$ & $A_1$ & & &\\ \hline
$d=4$ & $E_7$ & $D_6$ & $A_5$ & $A_1\times A_3$ &
        $\R \times A_2$ & $\R \times A_1$ {\it or} $A_2$& $\R$ &+ &  \\ \hline
$d=3$ & $E_8$ & $E_7$ & $E_6$ & $E_5$ & $E_4$ & $A_1\times A_2$ & $\R\times
A_1$ & $\R {\it or} A_1$&+\\ \hline
\end{tabular}
\end{center}
\caption{The split magic triangle \cite{pbl}.}
\end{sidewaystable}

\subsection{Del Pezzo surfaces and coset theories}

Let us first say a few words on del Pezzo surfaces and how they are related to
physical theories. For a detailed exposition and mathematical background,
see \cite{pbl}.

A (complex) del Pezzo surface is an algebraic, projective, complex surface
which has an ample anticanonical divisor. A complex surface (which has
complex dimension two) is said to be projective if it can be embedded in
some $\CP^n$ in such a way that, at least locally, it is defined as the
zero locus of some homogeneous polynomials. The maximal exterior power of the
tangent bundle is a line bundle, called the anticanonical divisor and
denoted $-K$. The
property of being ample is that its first Chern class is positive : it gives
a positive integer when integrated on any 1-cycle.

Smooth del Pezzo surfaces are $\CP^2$, $\CP^1 \times \CP^1$ and $\CP^2_{r}$
blown up in $r \leq 8$ points in general position. Blowing up a point means
replacing it by the $\CP^1$ of all tangent directions.

The $H_2$ lattice of integral cohomology (or Picard group) of a del Pezzo surface is endowed with 
an intersection
product. We recall that the Picard
group of $\CP^2_{r}$ is generated by the
projective line $H$ of $\CP^2$ and the $r$
exceptional curves $E_i$ and the Picard group of $\CP^1 \times \CP^1$ is generated by
the two lines $l_1$ and $l_2$ satisfying $l_1^2=l_2^2=0$ and
$l_1.l_2=1$.

One can construct Du Val singularities by contracting chains of
curves of self-intersection -2. The type of the singularity depends on the
intersections of the (-2)-curves considered and is described by a Dynkin
diagram. We will consider here $A_k$ singularities.

In \cite{inv} it was remarked that BPS states of M-theory compactified on
orthogonal tori correspond to divisors of vanishing (virtual) genus (called rational
divisors) on smooth del Pezzo's. In \cite{pbl} we have shown that one can
associate a Borcherds generalised Cartan matrix to these surfaces, from which we were 
able to recover
the equations of motion of all p-forms of the theories, and relations
between tensions of BPS objects. Moreover, we associated to each theory of
the oxidation triangle of \cite{cjlp} a del Pezzo surface which gives its
bosonic field content and equations of motion. The symmetry of the triangle,
on the del Pezzo side of the correspondence, is the fact that each pair of
symmetric del
Pezzo's have the same degree zero divisors. We now give
the precise proof of this fact, already sketched in \cite{pbl}.

\subsection{Proof of the triangle symmetry}

Let us first recall that for del Pezzo surfaces, the symmetry of
the SM$\Delta$ follows from the
fact that a del Pezzo surface of degree $n+1$ ($K^2=n+1$ with $K$ the
canonical divisor) with
exactly one singularity of type $A_k$ has the same group of
reflections as the del Pezzo of degree $k+1$ and singularity
$A_n$. This group  is the group of reflections with respect to the
(-2)-divisors orthogonal to $K$. In fact, one can check that the
sublattices of the Picard group orthogonal to $K$, ($K^\perp$), are 
identical for both cases and not only the (-2)-divisors.

Let us now show that given any del Pezzo surface, at least when
it is normal, we can construct a symmetric triangle such that this
surface lies on the first column of it.  Let us start with some given
singularities on this surface, in the following we shall focus on one
additional singularity. When the starting surface is non-singular
we recover the case of the previous paragraph. We may first blow up
points in general position in order to get a del Pezzo surface
with $K^2=1$. We claim that if we can blow down $n$
non-intersecting (-1)-curves isomorphic to $\CP^1$ on that new surface
to get a del Pezzo of degree $n+1$, we can alternatively construct
another del Pezzo of degree 1 with an additional singular point
$A_n$ with the same $K^\perp$, and conversely. Applying this
twice, we see that we can construct a del Pezzo surface of degree
$n+1$ and one singular point $A_k$ which has the same $K^\perp$ as
a del Pezzo of degree one and  two singular points $A_k$ and
$A_n$. The same reasoning tells us that this surface has the same
$K^\perp$ (full homologies may differ) as a third one of degree
$k+1$ and singularity $A_n$, which proves the symmetry of the
triangle of split U-dualities with respect to the diagonal.

Now, let us explain why for a del Pezzo surface of degree one
there is a one-to-one correspondence between blowing down  $k$
non-intersecting (-1)-curves isomorphic to $\CP^1$ and an $A_k$
singularity. At the level of the Picard group, blowing down curves
corresponds to keeping their orthogonal complement. Let $E_i$,
with $i$ running from 1 to $k$ be the so-called exceptional
(-1)-spheres. As they do not intersect, we have
$E_i.E_j=0$, and of course $E_i^2=-1$ and $K.E_i=-1$.
The vector space spanned
by these divisors is also generated by the divisors $E_i-E_{i+1}$
($1 \leq i \leq k-1$) and $-E_1$. So the orthogonal complement of
these two sets of divisors is the same. This is a result at the
level of divisors without assurance at this stage that the second
set can be realised effectively by spheres. If we restrict
ourselves to $K^\perp$ in the Picard group, we can also exchange
any $E_i$ with $-E_i-K$, and we see that inside
the subspace of the Picard group perpendicular to $K$, the
orthogonal complement of the $E_i$'s, with $1 \leq i \leq k$ is
also the orthogonal complement of $E_i-E_{i+1}$, with $1 \leq i
\leq k-1$ and $-E_1-K$ which are themselves (-2)-divisors  in
$K^\perp$. Moreover, one can check, keeping in mind $K^2=1$, that
these are virtual spheres in an $A_k$ configuration in other words
whose intersection matrix is the opposite of an $A_k$ Cartan
matrix.

One can easily see that this argument can be inverted
and that starting with such an $A_k$ of (-2)-divisors in the Picard
group, we can find $k$ orthogonal (-1)-divisors giving the same
orthogonal complement in $K^\perp$.
Now the key point is to decide which divisors are actually curves.

In order to construct an $A_k$ singularity from these (-2)-divisors, we
have to deform the del Pezzo surface such that these become
effective (-2)-spheres. Let us call $p_i$ and
$p_{i+1}$ the points we get by blowing down $E_i$ and $E_{i+1}$. For
$E_i-E_{i+1}$ to be a curve, we have to take $p_i$ and $p_{i+1}$
infinitely close, this preserves the orthogonality of $E_i$ and $E_{i+1}$
as
it is a continuous process. In other words, $p_{i+1}$ must lie on
$E_i$. If the del Pezzo surface is normal, once all singularities are
blown up, we get the smooth del Pezzo surface $\CP^2_{8}$, 
which corresponds to blowing up $\CP^2$ on eight almost general points, 
and the anticanonical divisor can be written as
$-K=3H-E_1-E_2'-\ldots-E_8'$ and therefore we have
$-K-E_1=3H-2E_1-E_2'-\ldots-E_8'$.
This divisor is a curve if the
eight points corresponding to the exceptional divisors $E_1, E_2',
\ldots E_8'$ lie on a cubic curve of $\CP^2$ which has a double point
that we take as $p_1$. Thus, we can construct an $A_k$ chain of
(-2)-curves to be blown down in order
to get the del Pezzo surface we wanted.

When the surface is nonnormal, the very last point is more subtle and
we do not know if such an $A_k$ singularity can always be constructed.
We will see below a case where it is possible, starting with the del
Pezzo surface we associated to a truncated version of Type I or
Heterotic String theory.

\subsection{Superalgebras}

On the superalgebra side of the dictionary, the translation is the
following. The symmetry of the split triangle comes from the fact that in
superalgebras corresponding to $\sigma$-models in $d=3$, the
centralizer of $n-1$ commuting $su(1|1)$ and one $sl(k+1)$ (also
commuting with the former)
has the same degree 0 part as
the centralizer of $k$ $sl(1|1)$ and one $sl(n)$, all commuting with
each other. In other words, instantons of both theories are the same,
and so is the U-duality algebra.
In the supersymmetric case the symmetry is only approximate. One should 
still factorise a split $sl(n)$ while going up any column ie under oxidation
this is related to the fact that this algebra is the remnant of diffeomorphisms.
But the rule we must adopt now for going from left to right at fixed dimension is 
rather to disintegrate the 
R-symmetry group, again starting from its affine root as in oxidation. However one 
is now factorising $ su(n)$'s.

\subsection{Split Triangle constructed from Type I}

In \cite{pbl} we showed that a del Pezzo surface can be associated to
Type I or Heterotic theories when one forgets the gauge sector. This
surface is not normal, which means that there is a singularity of
(co)dimension 1. We show here that a symmetric triangle of del Pezzo
surfaces can be constructed from it nevertheless, corresponding to
$\sigma$-models oxidations (Table 2).

The starting surface can be obtained in the following manner. We start
from $\CP^2$ and blow up one point, $p_1$, on it. Then we blow up a
second point, $p_2$, lying on the corresponding exceptional divisor
$E_1$ which gives another exceptional divisor $E_2$.
Then there is a (-2)-curve $E_1-E_2$, which is the transform of $E_1$
by the second blow-up, and a third (-1)-curve $H-E_1-E_2$ (where
$H$ is the line of $\CP^2$) that we blow down. This surface is a
$\CP^1$-bundle over $\CP^1$ and its Picard group
is generated by the section $E_1-E_2$ and the fiber
$H-E_1$. Now we blow up the intersecting point $p_3$ of $E_1-E_2$ with one
of the fibers, which gives an exceptional divisor $E_3$. The
(-2)-section becomes a (-3)-curve $E_1-E_2-E_3$. After blowing
down the (-1)-curve $H-E_1-E_3$, we get another $\CP^1$-bundle over
$\CP^1$ which still has fibers $A=H-E_1$ but which has now a
(-3)-section $E_1-E_2-E_3$. We repeat this operation once more and
get a bundle with the same fiber but with a (-4)-section $B=E_1-E_2-E_3-E_4$.

This (-4)-curve $B$ which is a conic can be projected to a double
line, and we get the surface described in \cite{pbl} with $-K=B+6A$, which
corresponds to the Type I theory (without gauge sector) in 10
dimensions. We get the surfaces corresponding to its toroidal
compactifications by blowing up points in general positions, as in
the regular case. The corresponding exceptional curves are noted $E_i'$. 
To construct a triangle on the right of this
column (Table 1), we have seen above that the divisor $-K'-E_1'$ must be an
irreducible curve, where $-K'=B+6A-E_1'-E_2'-\ldots-E_7'$ is the
anticanonical divisor of the $d=3$ surface, which
is obtained by blowing up seven points ($p_1'$ to $p_7'$) on the
$d=10$ one. As we have $-K'-E_1'=6H-5E_1-E_2-E_3-E_4-2E_1'-E_2'-
\ldots -E_7'$, we see that this is a curve if we take all the
blown-up points on a sextic such that $p_1$ is a quintuple point
and $p_1'$ a double point, which is possible. Indeed such a curve is
defined by an homogeneous polynomial of degree 6 in 3 variables, which
has 28 coefficients, which leaves 27 free coefficients after modding
out by the action of $\mathbb{C}^*$, and this is larger than 16, the
number of fixed points.

\begin{sidewaystable}[p]
\begin{center}
\begin{tabular}{|c||c|c|c|c|c|c|c|c|}
\hline
 &$k=1$&$k=2$&$k=3$&$k=4$&$k=5$&$k=6$&$k=7$&$k=8$\\
\hline
\hline
$d=10$ &$\R$ \\
\cline{1-2}
$d=9$ &$\R^2$ \\
\cline{1-3}
 $d=8$ & $\R \times A_1^2$ & $\R \times A_1$  \\
\cline{1-5}
 $d=7$ & $\R \times A_3$ & $\R^2 \times A_1$ & $\R^2$ & $\R$  \\
 \cline{1-6}
  $d=6$ & $\R \times D_4$ or $D_5$& $\R \times
A_1^3$ or $A_1\times A_3$ & $\R\times A_1^2$ & $\R^2$ or $A_1^2$ & $\R$   \\
\cline{1-6}
 $d=5$ & $\R \times D_5$ & $\R \times A_1 \times
A_3$ & $\R^2 \times A_1^2$ & $\R \times A_1^2$ & $\R^2$  \\
\cline{1-7}
 $d=4$ & $A_1 \times D_6$ & $D_6$ or $A_1^2
\times D_4$ & $\R \times A_1 \times A_3$ & $A_1\times A_3$ or $\R
\times A_1^3$& $\R^2 \times A_1$ & $\R \times A_1$ \\
 \cline{1-9}
$d=3$ & $D_8$ & $A_1 \times D_6$ & $\R \times D_5$ & $\R
\times D_4$ or $D_5$ & $\R \times A_3$ & $\R \times A_1^2$ & $\R^2$
& $\R$
\\
\cline{1-9}
\end{tabular}

\end{center}
\caption{Split Magic Type I-Heterotic Triangle}
\end{sidewaystable}

\bigskip
We observe that the del Pezzo surface ($k=1$, $d=6$) with Weyl group $D_5$ corresponds to type IIB
compactified on $T^4 / \Z_2$ without the twisted sector. We can obtain the other surface 
in the same slot (with Weyl group $D_4$) by applying
a birational transformation which corresponds to a blow up and a blow down of (-1)-curves. On the supergravity side, it corresponds to
a T-duality. The field contents of the various field theories obtained includes  a metric, 
a dilaton, a two form and some axions corresponding to the U-duality algebra.

\section{Real del Pezzo surfaces}

A {\it real algebraic variety} $X_\RR$ is
defined as a complex algebraic variety with an antiholomorphic involution
$\sigma_X$, called the {\it real structure} \cite{sih,kol1}. This real structure defines 
an involution of the Picard group
$Pic(X_\RR)$ in the following way: if $D$ is a Cartier divisor, it can be represented by 
meromorphic functions $f_i$ on an open cover ${U_i}$ of $X_\RR$ (with 
${f_i \over f_j}$ holomorphic and non-vanishing on $U_i \cap U_j$), then $\sigma(D)$ is 
given by
$(U_i,f_{i}^{\sigma_X}:=j \circ f_i \circ \sigma_X)$ where $j:\CC \rightarrow \CC$ is the 
complex conjugation.
The intersection of two divisors $D_1$ and $D_2$ defined as  
$D_1.D_2:=\int_{X_{\RR}} c_1(D_1)\cup c_1(D_2))$
(with $c_1$ the first Chern class) is then preserved by the real structure : 
$\sigma_X(D_1).\sigma_X(D_2)=D_1.D_2$.
Moreover, it can be proved that the anticanonical class $K_X$ is invariant under 
$\sigma_X$.
\\
We recall that if we blow up a point $P$ on a complex surface $X_{\CC}$,
we obtain a surface $Y_{\CC}$ and get a projection morphism $\pi: Y_{\CC} 
\rightarrow X_{\CC}$ such that $E=\pi^{-1}(P)$ is an exceptional curve of the first 
kind (that is $E
\simeq {\mathbb C}{\mathbb P}^1$ and $E.E=-1$) and 
$Y_{\CC} \setminus  \pi^{-1}(P) \simeq X_{\CC} \setminus  P$. The monoidal transformation 
$\pi$ is defined as follows:
let $(z_1,z_2)$ be local coordinates at $P$ defined in an open set $U$, then 
$V=\pi^{-1}(U)$ has equation $l_1z_2=l_2z_1$
in $U \times \CP^1$, where $(l_1,l_2)$ are homogenous coordinates of $\CP^1$. If we now 
blow up a point $P$ on a real surface $X_{\RR}$ with involution $\sigma_X$, we obtain a 
real surface $Y_{\RR}$ with a natural real structure $\sigma_Y$ defined by
$\sigma_Y(z_1,z_2,l_1,l_2)=(\sigma_X(z_1),\sigma_X(z_2),\bar{l}_1,\bar{l}_2)$ on $V$ where 
$\bar{}$ denotes complex
conjugation in $\CP^1$. This involution coincides with the one induced on 
$Y_{\RR} \setminus  \pi^{-1}(P)$ by identification with $X_{\RR} \setminus  P$.
Then, if $P$ is a real point, we obtain
an exceptional curve of the first kind $E$ satisfying $\sigma_Y(E)=E$ and for
complex conjugate points $P_1$ and $P_2$, we have
$\sigma_Y(E_1)=E_2$ and $E_1.E_2=0$ as $P_1$ and $P_2$ are distinct
points. Conversely, if $E$ is such a real (resp. a pair of complex 
conjugate) exceptional curve(s), $Y_{\RR}$ can be blown down to a surface
$X_{\RR}$ such that $E$ (resp. $E$ and $\sigma(E)$) is contracted to a
smooth real (resp. complex conjugate) point(s) $P$ and $Y_{\RR}$ is
precisely the blow-up of $X_{\RR}$ at the point $P$ as described above.
\\

A {\it real del Pezzo surface} $X_\RR$ is by definition a connected surface with
a real structure $\sigma$ that is possibly singular but Gorenstein
and is such that $-K$ is ample. In the complex case, we have already seen that the smooth
del Pezzo surface $X_{\CC}$ are $\CP^1 \times \CP^1$ and $\CP^2_{r}$ obtained by
blowing up $r$ points in general positions on the projective
surface $\CP^2$. The roots of $Pic(X_{\CC})$, defined as rational divisors orthogonal to $-K$, 
form a Dynkin diagram and generate a Weyl group $W$. It can be proved that
 $W$ is the group of isometries of  $Pic(X_{\CC})$, leaving $K$ fixed \cite{man, dem}.
Then, the real structure $\sigma$ on $X_\RR$ belongs to a conjugacy class of $W$. We use 
$\sigma$ to decompose the root lattice $L$ into eigenspaces $L^+ \oplus L^-$. It is known
that $\sigma$ is a product of dim $L^-$ reflections and the subgroup $W_{\sigma}$ of $W$ 
leaving $L^+$ pointwise fixed is itself a group generated by reflexions. We write 
$\triangle(\sigma)$ for its Dynkin diagram. The real structure $\sigma$ is entirely 
characterized by $\triangle(\sigma)$ (Table 2) \cite{wal1}.
We note that if we start with a real
del Pezzo surface $(X_{\RR}^r , \sigma)$ and blow up one real point $P$
(resp. complex conjugate points $P_1$ and $P_2$), we obtain a real del Pezzo
surface $(X^{r+1}_{\RR} , \sigma ^{'})$ with $\triangle(\sigma ^{'}) =
\triangle(\sigma)$ (resp. $\triangle(\sigma ^{'})=\triangle(\sigma)
\times A_1$ whose new root $E_1-E_2$ is anti-invariant under $\sigma ^{'}$).

We can now write down explicitly the involutions appearing in Table 3. We have already seen 
that the involution $A_1^r$ correspond to blow up
$r$ pairs of complex conjugate points.
If $d=8$, $A_1^{'}$ corresponds to
$\sigma_{\tiny A_1^{'}}(E_i)=E_i+E_0-E_1-E_2-E_3$. $A_1^{2^{'}}$ ($d=6$) and
$A_1^{3^{'}}$ ($d=4$) are now obtained from $A_1^{'}$ by blowing
up one or two pairs of complex conjugate points. The involution $A_1^{'}$ can be put
in a more standard way by blowing down the $(-1)$-curves
$H-E_1-E_2$ and $E_3$ exchanged by $\sigma_{\tiny A_1^{'}}$ . We obtain a del Pezzo
surface whose Picard group is generated by the lines $l_1:=H-E_1$ and $l_2:=H-E_2$, 
it is invariant under
$\sigma_{\tiny A_1^{'}}$ . This surface corresponds to the real form of the del Pezzo 
surface $\CP^1 \times \CP^1$  with the trivial involution $1$. It is a quadric hypersurface
$x_1^2+x_2^2-x_3^2-x_4^2=0$, denoted $Q^{(2,2)}$. So, the involution $A_1'$ can be obtained
by blowing up two complex conjugate points on $Q^{(2,2)}$. If we blow up a real point on 
$Q^{(2,2)}$, we obtain
a surface isomorphism to $\CP^2$ with the trivial involution and blown up on two real points.
A second real form of the del Pezzo surface $\CP^1 \times \CP^1$
can be obtained by imposing an $A_1$ involution given by $\sigma_{A_1}(l_1)=l_2$ and we 
have the
quadric hypersurface $x_1^2+x_2^2+x_3^2-x_4^2=0$, denoted
$Q^{(3,1)}$. This surface, blown up on a real point, is isomorphic to $\CP^2$ with the 
trivial involution and blown up on two
complex conjugate points.

The involution $D_{2s}$ is
\begin{eqnarray}
\sigma(H)&=&(s-1)(H-E_1)-K-H \\
\sigma(E_1)&=&-E_1-K+(s-3)(H-E_1) \\
\sigma(E_i)&=&H-E_1-E_i \, \, \, \, (2 \leq i \leq 2s+1)
\end{eqnarray}
The involution $E_7$ (resp. $E_8$) correspond to the Geiser (resp.
Bertini)
involution and acts on $Pic(\CP^2_{7})$ (resp. $Pic(\CP^2_{8}))$ as $\sigma(D)=(D.K)K-D$ 
(resp.
$\sigma(D)=2(D.K)K-D$).

\begin{sidewaystable}[p]
\begin{center}
\begin{tabular}{|c||c|c|c|c|c|c|c|c|c|c|c|c|}

\cline{1-2}
 $d=K^2+2$ & $Weyl({\mathbb C}{\mathbb P}^2_{11-d})$  \\

\cline{1-3}
  $11$ & $+$ & $1$ \\
\cline{1-3}
  $10$ & $+$ & $1$ \\
\cline{1-4}
  $9$ & $A_1$ & $1$& $A_1$ \\
\cline{1-5} \cline{7-7} \cline{8-8}
  $8$ & $A_1 \times A_2$ & $1$ &$A_1$ &&& $A_1^{'}$& $D_2 \simeq A_1^2$ \\
\cline{1-5} \cline{7-7} \cline{8-8}
  $7$ & $A_4$ & $1$ & $A_1$ &$A_1^2$ \\
\cline{1-5}\cline{7-9}
  $6$ & $D_5$ & $1$ & $A_1$ &$A_1^2$ &&$A_1^{2^{'}}$ &$D_2 \times A_1$ &$D_4$ \\
\cline{1-9}\cline{8-9}
  $5$ & $E_6$ & $1$& $A_1$&$A_1^2$& $A_1^3$&&&$D_4$ \\
\cline{1-12}
  $4$ & $E_7$ & $1$ & $A_1$ & $A_1^2$ & $A_1^3$ & $A_1^3{'}$ & $D_2 \times
  A_1^2$ &$D_4$ &
 $D_4
\times A_1$ & $D_6$ & $E_7$  \\
\hline
 $3$  & $E_8$ & $1$ & $A_1$ & $A_1^2$ & $A_1^3$ & $$ &  $A_1^4$ & $D_4$&$D_4 \times
A_1$ & $D_6$ & $E_7$ & $E_8$ \\
\hline
\end{tabular}
\end{center}
\caption{Conjugacy classes of involutions, $\triangle(\sigma)$ Weyl subgroups of $\CP^2_{11-d}$ }

\end{sidewaystable}

\section{Real forms of Lie algebras}
In this section, we will explain briefly how to classify the {\it real Lie
algebras}
according to the {\it Tits-Satake diagram}. 
A real form of a complex Lie algebra $\G^{\CC}$ is defined by the restriction
 of the field of coefficients from the
complex to the real numbers.
In M-theory, we must eventually replace the real field by the integers.
For example, the choice of real coefficients in the Cartan-Weyl
basis defines the split form and its existence follows from the fact that the 
structure constants of $\G^{\CC}$
can be taken to be integers. For $SL(2,\CC)$, the split form is $SL(2,\RR)$ wheras the 
compact form is
$SU(2)$ and the arithmetic group is the famous modular group $SL(2,\Z)$. These diagrams 
of the real forms of
the classical and exceptional complex groups have been classified by Araki \cite{ara}.

Let $\G$ be a real simple Lie algebra with Cartan decomposition $\G =
\T \oplus^\bot  \P$ with $\T$ a maximal compact subalgebra of $\G$ and $\P$ the
orthogonal complement of $\T$ with respect to the Killing form which is negative definite 
when restricted to $\T$.
 We can define a linear Cartan involution
$\theta$ by $\theta\mid_{\P}=-1$ and $\theta\mid_{\T}=1$. For the split form, the 
Cartan-Chevalley involution $\theta$ is given by $\theta(h_\alpha)=-h_\alpha$, 
$\theta(e_\alpha)=-e_{-\alpha}$. Then, the maximal compact subalgebra $\T$ is generated
over $\RR$ by $\{e_\alpha-e_{-\alpha}\}$ and $\P$ by $\{h_\alpha\}$ and 
$\{e_\alpha+e_{-\alpha}\}$.

Let $\A$ be a maximal abelian subalgebra of $\P$, and let $\eta$ be (a
maximally
non compact $\theta$-stable)
Cartan subalgebra containing $\A$ ( $\A^*$ is its dual) . The dimension of $\A$ corresponds 
to the real rank of $\G$.
For the split form, $\A$ is generated over $\RR$ by the Cartan generators 
$\{h_\alpha\}$ 
and the real rank coincides with the rank of its complexification.
The conjugation $\sigma$ of $\G^{\CC}$, the complexification of $\G$, with
respect
 to $\G$ defines an involutive automorphism and all the fixed elements of
$\G^{\CC}$
by $\sigma$ are the real form $\G$.
Thus, we have $\sigma(X)=X$ and $\sigma(iX)=-iX  \, \, \forall X \in \, \, \G$.
In the
same way, we can define
the conjugation $\tau=\theta.\sigma$ of $K\G^{\CC}$ with respect to the
compact
real form $\G=\T \oplus i\P$.
The Satake
diagram permits to obtain
easily  the conjugation $\sigma$
and then to find the real form associated to it.

The Satake diagram of $(\G , \A)$ consists of

1.  The Dynkin diagram of $(\G^{\CC},\eta^{\CC})$. The root lattice is noted $\Delta$.

2.  A coloring of the vertices of the diagram: black if the
associated simple root of $(\G^{\CC},\eta^{\CC})$ restricts to $0$ on $\A^*$, white
otherwise. Let $\bar{\alpha}={1 \over 2}(\alpha-\theta(\alpha))$
denote the restriction of $\alpha \in \Delta$ to the subspace
$\A^*$ and $m_{\psi}=dim \{ \alpha \in \Delta \, \mid \, \bar{\alpha}=\psi \}$, called the 
multiplicity of $\psi$. We set $\Delta_0 = \{ \alpha \in \Delta : \bar{\alpha}=0 \}$
and $\Delta_1=\Delta - \Delta_0$.

3.  a "curved arrow" joining
two white vertices if and only if the associated simple roots $\alpha$ and $\beta$ 
restrict to
the same root on $\A^*$: $\bar{\alpha}=\bar{\beta}$.

4.  The Dynkin diagram of the root system: $\Sigma=\{\bar{\alpha}: \alpha \in \Delta_1
\}$.

A table of Satake diagrams of the real  simple Lie
algebras can be found in Helgason \cite{hel}. The involution $\sigma$ can
be
determined by the Satake diagram of $\G$. Indeed, it can be proved that we
can
choose a basis of $\eta^{\CC}$ such as
$\tau(\alpha)=-\alpha$ and so $\sigma=-\theta$ on the root lattice.
So, if the vertex $\alpha_i$ in the
Satake diagram of $\G$ is black, the action of $\sigma$ is
$\sigma(\alpha_i)=-\alpha_i$ and
if two white vertices $\alpha_i$ and $\alpha_j$ are joined by a curved arrow,
we have $\sigma(\alpha_i)+\alpha_i=\sigma(\alpha_j)+\alpha_j$. Then, the action of
$\sigma$
on the other roots can be deduced by using the facts that $\sigma$ is an
involution
which must preserve the Cartan matrix. More precisely, let $\Delta_1=\{\alpha_1,\cdots, 
\alpha_r\}$ and $\Delta_0=\{\alpha_{(r+1)},\cdots, \alpha_s\}$, then 
$\sigma(\alpha_i)=\alpha_{\pi(i)}+\sum_{i=r+1}^{s}c_{ij}\alpha_j$ where $\pi$ is an 
involutive permutation of $\{ 1,2,\cdots,r \}$ and $c_{ij}$ are non-negative integers. 
The proof is very simple: let us decompose
$\sigma (\alpha_i) =\sum m_{ij} \alpha_j+\sum_{i=r+1}^{s} c_{ij}\alpha_j$ on the basis 
$\Delta$. Now applying $\sigma$ to this identity, we have 
$\sum_{j} m_{ij}m_{jk}=\delta_{ik}$ which means that $m_{ij}$ can be considered as the elements of the permutation matrix and we obtain the desired result.
  For example, if $\G $ is the split
form, $\sigma$ is the identity. For the non-split $E_{{7}|{-5}}$ of real rank $4$, we give 
as an exercice
the involution $\sigma$ (Table 4). The associated restricted root system of $\Sigma$ is that 
of its maximal split subalgebra $F_4$ \cite{hj0}.

\begin{table}[!h]
\begin{center}
\begin{tabular}{|c|c|c|}
\hline

$\Delta$ & $\Sigma$ &$\sigma$ \\
\hline
\hspace{5mm}
{
\begin{picture}(130,30)
\multiput(0,0)(20,0){6}{\circle{10}}
\put(60,20){\circle*{10}}
\put(0,-15){\makebox(0.4,0.6){ $\alpha_6$}}
\put(20,-15){\makebox(0.4,0.6){$\alpha_5$}}
\put(40,-15){\makebox(0.4,0.6){ $\alpha_4$}}
\put(60,-15){\makebox(0.4,0.6){$\alpha_3$}}
\put(80,-15){\makebox(0.4,0.6){ $\alpha_2$}}
\put(100,-15){\makebox(0.4,0.6){ $\alpha_1$}}
\put(0,0){\circle*{10}}
\put(40,0){\circle*{10}}
\put(60,30){\makebox(0.4,0.6){ $\alpha_0$}}
\put(5,0){\line(1,0){10}}
\put(25,0){\line(1,0){10}}
\put(45,0){\line(1,0){10}}
\put(65,0){\line(1,0){10}}
\put(85,0){\line(1,0){10}}
\put(60,5){\line(0,1){10}}
\end{picture}} &
{\small \begin{picture}(70,30)
\multiput(5,0)(20,0){4}{\circle{10}}
\put(5,-15){\makebox(0.4,0.6){ $\bar{\alpha}_5$}}
\put(25,-15){\makebox(0.4,0.6){$\bar{\alpha}_3$}}
\put(45,-15){\makebox(0.4,0.6){ $\bar{\alpha}_2$}}
\put(65,-15){\makebox(0.4,0.6){$\bar{\alpha}_1$}}
\put(45,40){\makebox(0.4,0.6){$m(2\bar{\alpha}_i)=1$}}
\put(45,20){\makebox(0.4,0.6){$m(\bar{\alpha}_i)=1$}}
\put(10,0){\line(1,0){10}}
\put(30,2){\line(1,0){10}}
\put(30,-2){\line(1,0){10}}
\put(37,0){\line(-1,2){5}}
\put(37,0){\line(-1,-2){5}}
\put(50,0){\line(1,0){10}}

\end{picture}}
&
$\begin{array}{cc}
\sigma( \alpha_0 )=& - \alpha_0 \\
\sigma(\alpha_1)=& \alpha_1  \\
\sigma(\alpha_2)=& \alpha_2  \\
\sigma(\alpha_3)=& \alpha_3+\alpha_0+\alpha_4 \\
\sigma(\alpha_4)=&-\alpha_4 \\
\sigma(\alpha_5)=& \alpha_5+ \alpha_4 + \alpha_6\\
\sigma(\alpha_6)=&-\alpha_6\\
\end{array}$
\\
\hline
\end{tabular}
\end{center}
\caption{Satake diagram of $E_{{7}|{-5}}$ }
\end{table}

\section{Real Magic triangle/square}

\subsection{Origins}

In \cite{jul1}, the {\it oxidation endpoint} of the pure
supergravity with $\N$ supersymmetries in four dimensions was
obtained. By the oxidation endpoint, we mean the supergravity
theory in the highest possible dimension whose toroidal
dimensional reduction gives back precisely to the pure
supergravity in four dimensions with $\N$ SUSY spinors. For example, the
pure supergravity $\N =8$ in $d=4$ can be obtained by dimensional
reduction of eleven-dimensional supergravity \cite{jul2}. The
oxidation sequence is presented in Table 5
and we have indicated the corresponding U-duality cosets
${G/K_G}\mid_{dim \, K_G}^{dim \, G}$. $G$ is a real Lie group and
$K_G$, which corresponds to the R-symmetry, is the maximal compact
Lie group of $G$. A similar magic triangle (Table 6), corresponding
to Maxwell-Einstein supergravities $\N =2$ $d=5$, has be obtained by
\cite{gun}. The following U-duality cosets can be guessed simply
by checking that the dimension of the cosets equals the number of
scalars. We recall that in order to obtain these U-duality groups,
some forms must be dualized. In the following section, we will
double all field strengths with their duals and we will infer an
even larger symmetry, which corresponds to a {\it  real form of a
Borcherds superalgebra}. 
The SM$\triangle$ and M$\triangle$ have  also recently been studied \cite{keu1}, 
\cite{keu2} using group theory. 

\begin{sidewaystable}[p]
\begin{center}
\begin{tabular}{|c||c|c|c|c|c|c|c|c|}
 \hline
$$&$\N =7$&$\N =6$&$\N =5$&$\N =4$&$\N =3$&$\N =2$&$\N =1$ & $\N =0$\\
\hline
$d=11$ & $+$ \\
\cline{1-2}
 $d=10$ &$\R\mid_{0}^{1}$ \\
\cline{1-2} $d=9$ & ${SL(2) \over SO(2)} \times \R \mid_{1}^{4}$  \\
\cline{1-2} $d=8$ & ${SL(3) \times SL(2) \over SO(3) \times
SO(2)}\mid_{4}^{11}$
 \\
\cline{1-2} $d=7$ & ${SL(5) \over SO(5)}\mid_{10}^{24}$\\
\cline{1-3} $d=6$ & ${SO(5,5) \over SO(5) \times
SO(5)}\mid_{20}^{45}$ & ${SO(5,1) \times SO(3) \over SO(5) \times
SO(3)}\mid_{13}^{18}$  \\
         \cline{1-3}
$d=5$ & ${E_6 \over Usp(8)}\mid_{36}^{78}$ & ${SU^\star(6) \over
Usp(6) }\mid_{21}^{35}$   \\
       \hline
$d=4$ & ${E_7 \over SU(8)}\mid_{63}^{133}$ & ${SO^\star(12) \over
U(6) }\mid_{36}^{66}$ & ${SU(5,1) \over U(5) }\mid_{25}^{35}$ &
${SU(4) \times SU(1,1) \over SU(4) \times SO(2) }\mid_{16}^{18}$ &
${U(3) \over U(3)}$ & ${U(2) \over U(2)}$ & ${U(1) \over U(1)}$
& +   \\
 \hline
$d=3$ & ${E_8 \over SO(16)}\mid_{120}^{248}$ & ${E_7 \over S0(12)
\times S0(3) }\mid_{69}^{133}$ & ${E_6 \over SO(10) \times SO(2)
}\mid_{46}^{78}$ & ${SO(8,2)  \over SO(8)
 \times SO(2)}\mid_{29}^{45}$ & ${SU(4,1)  \over SO(6) \times SO(2)}\mid_{24}^{16}$
        & ${SU(2,1) \times SU(2) \over SO(4)\times SO(2)}\mid_{7}^{11}$
        & ${SL(2) \times SO(2) \over SO(2) \times SO(2)}\mid_{2}^{4}$
& ${SL(2) \over SO(2)}\mid_{1}^{3}$ \\
\hline
\end{tabular}
\end{center}
\caption{Real magic triangle Cosets}
\end{sidewaystable}

\begin{table}[!h]
\begin{center}
\begin{tabular}{|c||c|c|c|}
\hline
$d=6$&${SO(9,1) \over SO(9)}\mid_{37}^{45}$&${SO(5,1) \times SO(3) \over SO(5) \times
SO(3)}\mid_{13}^{18}$ & ${SL(2,\CC)\times U(1) \over SU(2) \times
U(1)}\mid_{4}^{8}$  \\
\hline
 $d=5$ &${E_6 \over F_4}\mid_{52}^{78}$ & ${SU^{*}(6) \over Usp(6)}\mid_{21}^{35}$ & 
${SL(3,\CC) \over SU(3)}\mid_{8}^{17}$ \\
\hline
 $d=4$ &${E_7 \over E_6 \times U(1)}\mid_{69}^{133}$ & ${SO^{*}(12) \over
U(6)}\mid_{36}^{66}$ & ${SU(3,3) \over S(U(3) \times U(3))}
\mid_{17}^{35}$ \\
\hline
 $d=3$ &${E_8 \over E_7 \times SU(2)}\mid_{136}^{248}$ & ${E_7 \over
SO(12) \times SU(2)}\mid_{79}^{133}$ & ${E_6 \over SU(6) \times SU(2)}\mid_{38}^{78}$ \\
\hline
\end{tabular}
\end{center}
\caption{Magic square}
\end{table}

The oxidation sequence of each column $\N$ can be obtained by deleting a line of white
 roots  at the end of 
the extended Dynkin-Satake diagrams where the affine vertex attaches \cite{julia1}. 
We already mentioned the mysterious mechanism of disintegration of the R-symmetry that starts
also from the affine end but of the maximal compact U-duality Dynkin diagrams. Alternatively 
we can obtain the non-split U-duality algebras of each column of the Magic Triangle starting 
from the {\it Vogan diagram} of $E_{n|n}$. The Vogan diagrams encode the 
real forms of Lie algebras using a
maximal compact Cartan subalgebra \cite{knap}. The Vogan bicoloured diagrams may be chosen 
to have all or all but one compact vertices, 
represented now by white dots. Let us take the example of $n=8$. The white
dots with the affine root generate the Dynkin diagram of the maximal compact 
algebra $SO(16)$ and
the SUSY breaking sequence in $d=3$ amounts to the deletion of white dots starting from 
the affine root of $E_8$.

\subsection{Construction with real singularities}

We proved \cite{pbl} that the field theories of the SM$\bigtriangleup $
correspond to  normal del Pezzo surfaces $X_\CC$ with Du
Val singularities resulting from the contraction of a set of intersecting $(-2)$-curves 
forming a Dynkin diagram of type $A_*$ ie $Sl(n)$. In order to obtain other real forms of the 
U-duality groups than the split ones, it seems natural to
guess that  the supergravity theories of the original M$\bigtriangleup$ correspond to  real 
normal del Pezzo surfaces
$X_\RR$ with real (Du Val-type)
singularities which are associated to  real forms of $A_{*}$.  
The real forms of $A_{*}$ that
can be obtained are $Sl(n)$, $SU(n+1,n-1)$, $SU(n,n)$ and $SU(n,n-1)$ with
the Tits-Satake diagrams \cite{slo, wal2}  of Table 7.

\begin{table}[!h]
\begin{center}
\begin{tabular}{|c|c|}
\hline
$SL(n)$
& \begin{picture}(68,10)
\thicklines
\multiput(4,3)(30,0){3}{\circle{8}}
\put(8,3){\line(1,0){22}}
\multiput(38,3)(6,0){4}{\line(1,0){4}}
\end{picture}
\\
\hline
$SU(n+1,n-1)$ $n\ge 1$
& \begin{picture}(98,40)
\thicklines
\multiput(34,3)(0,30){2}{
\multiput(0,0)(30,0){3}{\circle{8}}
\put(4,0){\line(1,0){22}}
\multiput(34,0)(6,0){4}{\line(1,0){4}}}
\put(4,18){\circle*{8}}
\put(6,20){\line(2,1){24}}
\put(6,16){\line(2,-1){24}}
\multiput(34,18)(30,0){3}{\psline{<->}(0,-.3)(0,.3)}
\end{picture}
\\
\hline
$SU(n,n)$ $n\ge 1$
& \begin{picture}(98,40)
\thicklines
\multiput(34,3)(0,30){2}{
\multiput(0,0)(30,0){3}{\circle{8}}
\put(4,0){\line(1,0){22}}
\multiput(34,0)(6,0){4}{\line(1,0){4}}
}
\put(4,18){\circle{8}}
\put(6,20){\line(2,1){24}}
\put(6,16){\line(2,-1){24}}
\multiput(34,18)(30,0){3}{\psline{<->}(0,-.3)(0,.3)}
\end{picture}
\\
\hline
$SU(n,n-1)$ $n\ge 1$
& \begin{picture}(98,40)
\thicklines
\multiput(34,3)(0,30){2}{
\multiput(0,0)(30,0){3}{\circle{8}}
\put(4,0){\line(1,0){22}}
\multiput(34,0)(6,0){4}{\line(1,0){4}}
}

\put(18,18){\line(1,1){12}}
\put(18,18){\line(1,-1){12}}
\multiput(34,18)(30,0){3}{
\psline{<->}(-.1,-.3)(-.1,.3)
}

\end{picture}
\\
\hline
\end{tabular}
\end{center}
\caption{Satake diagrams of $A_{n-1}$ }
\end{table}

Then, the supergravity theory $(\N,d)$ of M$\bigtriangleup$ in
$d$ dimensions with $\N$ SUSY is identified with the normal real
del Pezzo surface $X_{\RR}$ with $d=K^2+2$ and a real
singularity $A_{(8-\N)}$ in the following way: the Satake
involution which acts on the root lattice $L$ of the Dynkin
diagram of the U-duality group $G$ is identified with the real
structure $\sigma$ of the real surface $X_{\RR}$. By using the
facts that $Pic(X_{\RR})=L \oplus \Z K$ and $\sigma(K)=K$ we
deduce the action of the real structure $\sigma$ on the Picard
group of $X_{\RR}$. 

Now, the $A_{(8-\N)}$ singularity,
characterized by another Satake involution $\sigma_{A}$, remains to be
determined. The Picard group of the normalized real del Pezzo
surface $\tilde{X}_{\RR}$, which is isomorphic in the complex case
to $\CP^2_{11-d}$,  is given by $Pic(\tilde{X}_{\RR})=Pic(X_{\RR})
\oplus^{\bot} A_{(8-\N)}$. So the real structure of
$\tilde{X}_{\RR}$ is $\tilde{\sigma} = \sigma \oplus \sigma_A$ and
must correspond to those listed in Table 3. In some particular
cases, we will obtain more than one solution. In the following
table (Table 8), the real normal del Pezzo surface $X_{\RR}$,
corresponding to the supergravity $(\N,d)$ of the
M$\bigtriangleup$, is noted $(\Delta(\tilde{\sigma}),A)$ where
$A$ is a real form of the $A_{(8-\N)}$ singularity. A similar table
(Table 9) can be obtained for the supergravity theories of the
magic square. It should be noted that the link between the number
of singularities and the number of supersymmetries is not obvious
as each theory of the magic square corresponds to $\N =2$ or more.

\begin{sidewaystable}[p]
\begin{center}
{\small
\begin{tabular}{|c||c|c|c|c|c|c|c|c|c|}
\hline
$$&$\N =7$&$\N =6$&$\N =5$&$\N =4$&$\N =3$&$\N =2$&$\N =1$& $\N =0$\\
\hline \hline
$d=11$ & + \\
\cline{1-2}
 $d=10$ &$(1,0)$ \\
\cline{1-2} $d=9$ & $(1,0)$ \\
\cline{1-2} $d=8$ & $(1,0)$ \\
\cline{1-2} $d=7$ & $(1,0)$ \\
\cline{1-3} $d=6$ & $(1,0)$ & $(A_1^3,SU(1,1))$
                       \\
  & & $(D_4,SU(1,1))$
                       \\
         \cline{1-3}
$d=5$                       & $(1,0)$ & $(A_1^3,SU(1,1))$  \\
                       & & $(D_4,SU(2))$  \\
       \hline
$d=4$ & $(1,0)$ & $(A_1^3,SU(1,1))$
& $(A_1 \times D_4,SU(2,1))$ & $(A_1 \times D_4,SU(3,1)) $ & $(A_1 \times D_4 ,SU(3,2))$ & $(A_1 \times D_4,SU(4,2))$ &
$(A_1^4,SU(4,3))$
   & $(1,SL(8))$\\
 & & $(D_4,SU(2))$
& & $( D_4,SU(2,2)) $ & & $(A_1^4,SU(3,3))$ &  & $(A_1^3,SU(4,4))$\\
& & & & & & & & $(D_4,SU(5,3))$\\
\hline
$d=3$ & $(1,0)$ & $(A_1^3,SU(1,1))$
& $(A_1 \times D_4,SU(2,1))$ & $(A_1 \times D_4,SU(3,1)) $ & $(A_1 \times D_4 ,SU(3,2))$ & $(A_1 \times D_4,SU(4,2))$ &
$(A_1^4,SU(4,3))$
   & $(1,SL(8))$\\
 & & $(D_4,SU(2))$
& & $( D_4,SU(2,2)) $ & & $(A_1^4,SU(3,3))$ &  & $(A_1^3,SU(4,4))$\\
& & & & & & & & $(D_4,SU(5,3))$\\
\hline
\end{tabular}}
\end{center}
\caption{Real magic del Pezzo's triangle : real structure on the resolution
(given by the anti-invariant subdiagram) and real form of the singularity.}
\end{sidewaystable}

\begin{table}[!h]
\begin{center}
\begin{tabular}{|c||c|c|c|}
\hline
 $d=5$ &$(D_4,0)$ & $(A_1^3,SL(2))$ & $(A_1^3,SU(2,1))$ \\
& & $(D_4,SU(2))$ & \\
\hline
 $d=4$ &$(D_4,0)$ & $(A_1^3,SL(2))$ & $(A_1^3,SU(2,1))$ \\
& & $(D_4,SU(2))$ & \\
\hline
 $d=3$ &$(D_4,0)$ & $(A_1^3,SL(2))$ & $(A_1^3,SU(2,1))$ \\
& & $(D_4,SU(2))$ & \\
\hline
\end{tabular}
\end{center}
\caption{Real magic del Pezzo's square }
\end{table}

We observe that, as in the complex case \cite{pbl}, each vertical
step down, which corresponds on the supergravity side to a
compactification on a circle, corresponds to blowing up one real point.
The singularity $A$, $\Delta{(\sigma)}$ and
$\Delta(\tilde{\sigma})$ are unchanged under this operation.

\subsection{Real Borcherds superalgebra}

We will show in this section how the connection between supergravity theories and real del Pezzo surfaces can be useful to show that the U-duality algebra can be enlarged into a real Borcherds superalgebra. The basic procedure is a direct extension of \cite{pbl}. According to this paper, we can associate to a complex del Pezzo surface $X_\CC$
 a Borcherds superalgebra whose simple roots $\alpha_i$ generate the full Picard group $Pic(X_\CC)$ and the set of positive roots contains the rational (i.e of vanishing virtual genus) 
divisor classes of non negative degree and also (except in the case of M-theory) the 
anticanonical class $-K$. 

We can then define minus the intersection matrix
$A_{ij}=-\alpha_i.\alpha_j$ and a $\Z_2$-graduation by $grad(\alpha_i)=-K.\alpha_i \, \, mod \, 2$. However, it turns out that whenever a fermionic root of square $-1$ appears it should be viewed as an $sl(1|1)$ superroot , i.e. it should have zero
Cartan-Killing norm.
The corresponding modified matrix will be our
{\it Cartan matrix} $a_{ij}$.
This matrix $a_{ij}$ satisfies the following properties
and thus defines a  Borcherds superalgebra (or {\it Generalized Kac-Moody}
  superalgebra).

\begin{eqnarray}
(i) & a_{ij} \leq 0 & {\mathrm if } \ i \neq j \\
(ii) & \frac{2a_{ij}}{a_{ii}} \in  {\mathbb Z} & {\mathrm if } \
a_{ii} > 0 \,\, , \,\, grad(\alpha_i)=0 \\
(iii) & \frac{a_{ij}}{a_{ii}} \in {\mathbb Z} & {\mathrm if } \
a_{ii} > 0 \,\, , \,\, grad(\alpha_i)=1  
\end{eqnarray}

The Borcherds superalgebra associated to the matrix $a_{ij}$
has a Cartan subalgebra $H$, with basis $\{h_{\alpha_i}\}$, it is  by
definition the Lie superalgebra $G$ generated by $H$ and by the
elements $e_{\alpha_i}$, $f_{\alpha_i}$ satisfying here the following
elementary relations and their consequences:

\begin{eqnarray}
(1) & [ e_{\alpha_i} ,f_{\alpha_j}  ] = \delta_{ij} h_{\alpha_i} \\
(2) & [ h_{\alpha_j} , e_{\alpha_i} ] = a_{ij} e_{\alpha_i}
\textrm{,} \,  [ h_{\alpha_j} ,
f_{\alpha_i} ] = -a_{ij} f_{\alpha_i} \\
(3) & [ h_{\alpha_i} , h_{\alpha_j} ] = 0 \\
(4) & {ad(e_{\alpha_i})}^{1-2\frac{a_{ij}}{a_{ii}}} e_{\alpha_j} = 0
= (ad(f_{\alpha_i}))^{1-2\frac{a_{ij}}{a_{ii}}} f_{\alpha_j}
\ \textrm{if} \ a_{ii} > 0 \\
(5) & [ e_{\alpha_i} , e_{\alpha_j} ] = 0 = [ f_{\alpha_i} ,f_{\alpha_j}  ]
\ \textrm{if} \ a_{ij} = 0
\end{eqnarray}
A Killing form is defined as $B(e_\alpha,e_{-\alpha})=-1$ and $B(h,h_\alpha)=\alpha(h)$ $\forall \, h \in H$.

Using this contruction, we can associate to a del Pezzo surface
$X_{\RR}$ with real structure $\sigma$ a real Borcherds superalgebra
$\G$ with a conjugation given by $\sigma$. However, it does not work
for all real structures on a given del Pezzo, because for real
superalgebras, one further imposes that the $\sigma$-system of roots satisfies the 
{\it normal condition} \cite{ara} which means that for any bosonic root $\alpha$, $\alpha-\sigma(\alpha)$ must not be a root. 
Following Araki, we can define {\it Satake superdiagrams} (appendix) which recently appeared 
for the classical Lie superalgebras case in \cite{pat}. All the real Borcherds superalgebras which will appear, have only fermionic imaginary simple roots, of vanishing norm, noted $\beta_i$. The other bosonic roots which generate the U-duality algebra will be noted $\alpha_i$. We note that the restricted roots $\Sigma$ form a Borcherds superalgebra $\bar{\G}$ and its Cartan matrix is noted $\bar{a}_{ij}$ .
Then, we color the superdiagrams as explained is section 4 (this convention differs from that of our previous paper). For example, the Borcherds superalgebra $B_{6}^6$ of the supergravity
$\N =6$ and $d=6$ is given in the following table by:

\vskip 2truemm
{\small
\begin{tabular}{|c|c|c|}
\hline

$\Delta$ & $\Sigma$ &$\sigma$ \\
\hline \hline
 \hspace{5mm} {\small \begin{picture}(120,60)
\multiput(0,40)(20,0){5}{\circle{10}}
\put(0,25){\makebox(0.4,0.6){ $\alpha_3$}}
\put(20,25){\makebox(0.4,0.6){$\alpha_2$}}
\put(40,25){\makebox(0.4,0.6){ $\alpha_1$}}
\put(60,25){\makebox(0.4,0.6){$\beta$}}
\put(80,25){\makebox(0.4,0.6){ $\alpha_0$}}
\put(0,40){\circle*{10}}
\put(40,40){\circle*{10}}
\put(80,40){\circle*{10}}
\put(5,40){\line(1,0){10}}
\put(25,40){\line(1,0){10}}
\put(45,40){\line(1,0){10}}
\put(65,40){\line(1,0){10}}
\end{picture}} &
{\small \begin{picture}(40,30)
\multiput(15,40)(20,0){2}{\circle{10}}
\put(15,20){\makebox(0.4,0.6){ $\bar{\alpha_2}$}}
\put(15,10){\makebox(0.4,0.6){ ${\bf 4}$}}
\put(35,20){\makebox(0.4,0.6){$\bar{\beta}$}}
\put(35,10){\makebox(0.4,0.6){ ${\bf 4}$}}
\put(20,40){\line(1,0){10}}
\end{picture}}
&
$\begin{array}{cc}
\sigma( \alpha_i )=& - \alpha_i \, \forall  i=0,1,2 \\
\sigma(\alpha_2)=& \alpha_2+\alpha_3+\alpha_1  \\
\sigma(\beta)=&\beta+\alpha_0+\alpha_1\\
\bar{a}=&\left(\matrix {2 & -1 \cr
                 -1 & -2}\right) \\

\end{array}$ \\

\hline
\end{tabular}}
\vskip 2truemm

In order to obtain  the equations of motion  of the supergravity theory and to show that 
they are invariant under the (truncated)
Borel supergroup of a real Borcherds superalgebra $\G$, we need to generalize the Iwasawa decomposition which can be studied in \cite{hel}.
The proof remains the same and the decompositon reads as $\G=\T+\A+\N$
where  $\N\equiv \sum_{\bar{\lambda} \in \Sigma^+} \G_{\bar{\lambda}}$ is the nilpotent subalgebra of positive restricted roots. The other notations are those used in section 2. We note that $dim(\G_{\bar{\lambda}})=m(\bar{\lambda}).mult(\bar{\lambda})$ where $mult(\bar{\lambda})$
is the mutiplicity of the $\bar{\lambda}$ for the Borcherds superalgebra $\bar{\G}$.
Let us introduce the following nonlinear "potential" differential form:
\begin{equation}
\v=\prod_{\bar{\lambda} \in \A^*} e^{h_{\bar{\lambda}} \phi_{\bar{\lambda}}}
\prod_{\bar{\lambda} \in \Sigma^+\, \mid \, d(\bar{\lambda}) \leq D }e^{e_{\bar{\lambda}} A_{\bar{\lambda}}}
\end{equation}
The Grassman real angle $A_{\bar{\lambda}}$ is a form of degree $d(\bar{\lambda}) \equiv -K.\bar{\lambda}$ lower than $d \equiv K^2+2$ ( $\phi_{\bar{\lambda}}$ is $0$-form) defined on a formal manifold of dimension $d$. The field strength
$\G \equiv d \v \v^{-1}$ is then a element of $\N$ whose coefficents are given by forms.
As in \cite{pbl}, we introduce the pseudo-involution $\S$ satisfying:
$$\S(e_{\bar{\lambda}})=e_{-\bar{\lambda}-K} \, \, \forall \bar{\lambda} \in \Sigma^+  \, \mid \, \, d(\bar{\lambda}) \leq d
$$
This operator is well defined because the truncated root lattice of the Borcherds 
superalgebra is invariant under this involution. Finally, if we combine the (twisted) 
self-duality equations $S\G=*\G$ and the Maurer-Cartan equation
$d\G=-\G\wedge\G$ which follows by taking the exterior derivative of the field strength, we obtain some second-order equations which reproduce the equations of motion for the $p$-forms 
of the supergravity theory with the corresponding real finite simple Lie algebra (\ie 
U-duality algebra) . These equations are trivially invariant under the Borel supergroup 
whose Lie superalgebra is generated by $\A+\N$. Indeed we can define a natural action on the 
generalized potential $\v$ by
$\v'=\v.\Lambda$ with $\Lambda$ an element of the Borel supergroup with coefficients given by closed forms. The field strength $\G$ is then invariant.
We have computed in the following table the positive restricted roots of $B_{6}^6$ with their multiplicities.
\begin{center}
\begin{tabular}{|c|c|c|c|}
\hline
Degree & positive root & mutiplicity & potential \\
\hline
0 & 0& &$\phi$\\
0 & $\bar{\alpha_2}$ &4 & $A_{(0)}$\\
1 & $\bar{\beta}$ & 4 & $A_{(1)}$\\
1 & $\bar{\alpha_2}+\bar{\beta}$  & 4 & ${\cal A}_{(1)}$\\
2 & $2\bar{\beta}$   & 1 & $A_{(2)}$\\
2 & $\bar{\alpha_2}+2\bar{\beta}$  & 1 & ${\cal A}_{(2)}$\\
2 & $2(\bar{\alpha_2}+\bar{\beta})$  &  1& $\tA_{(2)}$\\
3 & $\bar{\alpha_2}+3\bar{\beta}$  &  4 & ${\cal \tA}_{(3)}$ \\
3 & $2\bar{\alpha_2}+3\bar{\beta}$  & 4 & ${\tA}_{(3)}$ \\
4 & $\bar{\alpha_2}+4\bar{\beta}$   & 4 & ${A}_{(4)}$ \\
4 & $3\bar{\alpha_2}+4\bar{\beta}$   & 4 & none \\
4 & $2\bar{\alpha_2}+4\bar{\beta}$   & 1 & $-K$ \\
\hline
\end{tabular}
\end{center}

The root $3\bar{\alpha_2}+4\bar{\beta}$  is exchanged under the pseudo-involution $\S$ with the negative root $-\bar{\alpha_1}$, which corresponds to the NS-1 instanton which is S-dual to the D-1 instanton coupled to the scalar $A_{(0)}$. The root
  $3\bar{\alpha_2}+4\bar{\beta}$ is therefore the Hodge-dual of NS-1 and could be noted NS3.  We can check that we obtain
the right number of fields ie four axions, eight 1-forms, three 2-forms and the corresponding dual fields.

\section{Conclusion}
Recently, some other conjectured symmetries of M-theory such as Kac-Moody algebras $E_{10}$ 
and $E_{11}$, introduced a long time ago \cite{jul3} for the first one (and partly 
realised \cite{dam1}) and more recently \cite{carg} for the second one and further studied in
\cite{wes}  have reappeared. 
The hyperbolic Kac-Moody algebra $E_{10}$ appears when one studies the chaotic
behaviour of the M-theory near a cosmological singularity.
The solutions are chaotic if the corresponding Kac-Moody algebra is hyperbolic
which means that one obtains some finite dimensional 
simple Lie algebras or affine Lie algebras by
deleting a simple root to its Dynkin diagram \cite{dam2}. For the supergravity theory with 
$\N$ supersymmetries in  M$\triangle$, the corresponding over-extension can be found 
using the corresponding Borcherds superalgebra in $d=3$ dimensions. As noted in \cite{pbl},
 the fermionic simple root $\beta$ stands at the location of the affine root which is 
invariant 
under the Satake conjugation. The over-extension corresponds to add two invariant (white)
simple roots starting with   the affine root. On the del Pezzo side, it corresponds to 
blowing up two more real points. 

These algebras require also a new formulation of M-theory which incorporates the fields and 
their duals. However, in order 
to obtain these algebras, we need to dualize also the gravity field. Our supercoset 
construction based on some real Borcherds superalgebra must be extended in order to include 
these fields which can be described by multiforms \cite{hul}, \cite{bou}. For example, the 
gravity field and its dual are described by bi-forms. 
We are witnessing indeed magical coincidences with some defects that should be corrected to 
reach harmony and that in fact suggest a bigger structure yet to be discovered.
In particular the role of SUSY is still a puzzle.

\section{Appendix}
In this appendix, we list the real Borcherds superalgebra corresponding to each box of the 
SM$\Delta$.
We recall that $\beta_i$ is a fermionic root of vanishing norm and $\alpha_j$ is a bosonic 
root, normalized such that
${\alpha_{j}}^{2}=2$. 
$\bar{}$ corresponds to the projection under $\A$, the maximal  abelian subalgebra of $\P$ 
(cf section 4).
The bold number below a reduced simple root $\bar{\lambda}$ correspond to its multiplicity 
$dim(\G_{\bar{\lambda}})$ . 
We note that in $d=3$ dimensions, the fermionic simple root of the reduced Borcherds 
superalgebra stands at the location of the affine root. This is not so well defined for 
semi-simple non-simple Lie algebras. In fact  in order to obtain the 
right superalgebra for the supergravity theory $\N =2$ $d=4$ (resp. $\N =2$ $d=3$), we need 
to quotient the superBorcherds algebra by the following relation invariant under $\sigma$: 
$\alpha_0+\beta_2=\beta_1+\beta$ (resp. $\alpha_2+\alpha_1+\beta_1=\alpha_0+\beta_2$). 
Indeed, the dimension of the Picard group is of the corresponding real del Pezzo is $3$ 
(resp. $4$) and therefore, the roots are not independent. This is another instance of a 
feature that deserves further investigation \cite{pbl2}.

\begin{table}[htp]
\begin{center}
{\small
\begin{tabular}{|c|c||c|c|}
\hline

$\Delta$ & $\Sigma$ &$\Delta$ & $\Sigma$ \\
\hline {\small \begin{picture}(80,60)
\put(0,-10){
\multiput(0,40)(20,0){5}{\circle{10}}
\put(0,25){\makebox(0.4,0.6){ $\alpha_3$}}

\put(20,25){\makebox(0.4,0.6){$\alpha_2$}}
\put(40,25){\makebox(0.4,0.6){ $\alpha_1$}}
\put(60,25){\makebox(0.4,0.6){$\beta$}}
\put(80,25){\makebox(0.4,0.6){ $\alpha_0$}}
\put(0,40){\circle*{10}}
\put(40,40){\circle*{10}}
\put(80,40){\circle*{10}}
\put(5,40){\line(1,0){10}}
\put(25,40){\line(1,0){10}}
\put(45,40){\line(1,0){10}}
\put(65,40){\line(1,0){10}}
}
\end{picture}} &

{\small \begin{picture}(80,60)
\put(0,-3){
\multiput(15,35)(20,0){2}{\circle{10}}
\put(15,15){\makebox(0.4,0.6){ $\bar{\alpha}_2$}}

\put(50,55){\makebox(0.4,0.6){ ${\bf \N =6,d=6}$}}
\put(15,5){\makebox(0.4,0.6){ ${\bf 4}$}}
\put(35,15){\makebox(0.4,0.6){$\bar{\beta}$}}
\put(35,5){\makebox(0.4,0.6){ ${\bf 4}$}}
\put(20,35){\line(1,0){10}}
}
\end{picture}}
&

{\small \begin{picture}(70,60)
\put(0,-5){
\multiput(10,25)(20,0){2}{\circle{10}}
\multiput(10,45)(20,0){2}{\circle{10}}
\put(10,10){\makebox(0.4,0.6){ $\alpha_1$}}
\put(30,10){\makebox(0.4,0.6){ $\beta_1$}}

\put(10,55){\makebox(0.4,0.6){ $\alpha_2$}}
\put(30,55){\makebox(0.4,0.6){ $\beta_2$}}
 \put(10,25){\circle*{10}}
\put(10,45){\circle*{10}}

 \put(15,25){\line(1,0){10}}
\put(15,45){\line(1,0){10}}
 \put(10,30){\line(0,1){10}}
\put(30,30){\line(0,1){10}}

\put(37,35){\psline{<->}(0,-.3)(0,.3)}
}
\end{picture}} &

{\small \begin{picture}(80,60)
\put(0,-3){
\multiput(15,40)(20,0){1}{\circle{10}}

\put(50,55){\makebox(0.4,0.6){ ${\bf \N =3,d=4}$}}

\put(15,20){\makebox(0.4,0.6){ $\bar{\beta}_1$}}
\put(15,10){\makebox(0.4,0.6){ ${\bf 6}$}}
}
\end{picture}}
\\

\hline {\small \begin{picture}(82,60)
\put(0,-15){
\multiput(0,40)(20,0){5}{\circle{10}}
\put(0,25){\makebox(0.4,0.6){ $\alpha_3$}}
\put(20,25){\makebox(0.4,0.6){$\alpha_2$}}
\put(40,25){\makebox(0.4,0.6){ $\alpha_1$}}
\put(60,25){\makebox(0.4,0.6){$\alpha_{-1}$}}
\put(70,60){\makebox(0.4,0.6){$\beta$}}
\put(80,25){\makebox(0.4,0.6){ $\alpha_0$}}
\put(0,40){\circle*{10}} \put(60,60){\circle{10}}
\put(40,40){\circle*{10}} \put(80,40){\circle*{10}}
\put(5,40){\line(1,0){10}} \put(25,40){\line(1,0){10}}
\put(45,40){\line(1,0){10}} \put(65,40){\line(1,0){10}}
\put(60,45){\line(0,1){10}}
}
\end{picture}} &
{\small \begin{picture}(80,60)
\put(0,-8){
\multiput(15,40)(20,0){3}{\circle{10}}
\put(15,20){\makebox(0.4,0.6){ $\bar{\alpha}_2$}}
\put(15,10){\makebox(0.4,0.6){ ${\bf 4}$}}
\put(35,20){\makebox(0.4,0.6){$\bar{\alpha}_{-1}$}}
\put(35,10){\makebox(0.4,0.6){ ${\bf 4}$}}
\put(50,60){\makebox(0.4,0.6){ ${\bf \N =6,d=5}$}}

\put(55,20){\makebox(0.4,0.6){$\bar{\beta}$}}
\put(55,10){\makebox(0.4,0.6){${\bf 1}$}}
 \put(20,40){\line(1,0){10}}
\put(40,40){\line(1,0){10}}
}
\end{picture}}
& {\small \begin{picture}(70,60)
\put(0,-20){
\multiput(10,40)(20,0){2}{\circle{10}}
\multiput(10,60)(20,0){2}{\circle{10}}
\put(10,25){\makebox(0.4,0.6){ $\alpha_1$}}
\put(30,25){\makebox(0.4,0.6){ ${\alpha}_{-1}$}}
 \put(10,45){\line(0,1){10}}
\put(30,50){\psline{<->}(0,-.2)(0,.2)}
 \put(60,50){\makebox(0.4,0.6){$\beta$}}
  \put(10,70){\makebox(0.4,0.6){ ${\alpha}_2$}}
\put(30,70){\makebox(0.4,0.6){ $\alpha_{-2}$}}
 \put(10,40){\circle*{10}}
\put(10,60){\circle*{10}} \put(50,50){\circle{10}}
 \put(15,40){\line(1,0){10}}
\put(15,60){\line(1,0){10}}
 \put(35,40){\line(2,1){12}}
\put(35,60){\line(2,-1){12}}
}
\end{picture}} & {\small \begin{picture}(80,60)
\put(0,-8){
\multiput(15,40)(20,0){2}{\circle{10}}
\put(15,20){\makebox(0.4,0.6){ $\bar{\beta}_1$}}
\put(15,10){\makebox(0.4,0.6){ ${\bf 1}$}}
\put(50,60){\makebox(0.4,0.6){ ${\bf \N =3,d=3}$}}

\put(35,20){\makebox(0.4,0.6){ $\bar{\alpha}_{-1}$}}
\put(35,10){\makebox(0.4,0.6){ ${\bf 6}$}}

\put(20,40){\line(1,0){10}}
}
\end{picture}}
\\
\hline
{\small \begin{picture}(82,80)
\put(0,-15){
\multiput(0,40)(20,0){5}{\circle{10}}
\put(0,25){\makebox(0.4,0.6){ $\alpha_3$}}
\put(20,25){\makebox(0.4,0.6){$\alpha_2$}}
\put(40,25){\makebox(0.4,0.6){ $\alpha_1$}}
\put(60,25){\makebox(0.4,0.6){$\alpha_{-1}$}}
\put(77,60){\makebox(0.4,0.6){$\alpha_{-2}$}}
\put(80,25){\makebox(0.4,0.6){ $\alpha_0$}}
\put(72,80){\makebox(0.4,0.6){ $\beta$}} \put(0,40){\circle*{10}}
\put(60,60){\circle{10}} \put(40,40){\circle*{10}}
\put(80,40){\circle*{10}} \put(60,80){\circle{10}}
\put(5,40){\line(1,0){10}} \put(25,40){\line(1,0){10}}
\put(45,40){\line(1,0){10}} \put(65,40){\line(1,0){10}}
\put(60,45){\line(0,1){10}} \put(60,65){\line(0,1){10}}
}
\end{picture}} & {\small \begin{picture}(80,80)
\multiput(15,40)(20,0){4}{\circle{10}}
\put(15,20){\makebox(0.4,0.6){ $\bar{\alpha}_2$}}
\put(15,10){\makebox(0.4,0.6){ ${\bf 4}$}}
\put(35,20){\makebox(0.4,0.6){$\bar{\alpha}_{-1}$}}
\put(35,10){\makebox(0.4,0.6){ ${\bf 4}$}}
\put(50,70){\makebox(0.4,0.6){ ${\bf \N =6,d=4}$}}

\put(55,20){\makebox(0.4,0.6){$\bar{\alpha}_{-2}$}}
\put(55,10){\makebox(0.4,0.6){ ${\bf 1}$}}
\put(75,20){\makebox(0.4,0.6){$\bar{\beta}$}}
\put(75,10){\makebox(0.4,0.6){ ${\bf 1}$}}
\put(20,40){\line(1,0){10}} \put(40,38){\line(1,0){10}}
\put(40,42){\line(1,0){10}}
\put(43,40){\line(1,2){5}}
\put(43,40){\line(1,-2){5}}

 \put(60,40){\line(1,0){10}}
\end{picture}}
&{\small \begin{picture}(70,80)
\put(10,-10){
\multiput(0,40)(20,0){2}{\circle{10}}
\put(0,25){\makebox(0.4,0.6){ $\beta$}}
\put(20,25){\makebox(0.4,0.6){ $\beta_2$}}
\put(0,60){\circle*{10}}
\put(20,60){\circle{10}}
\put(0,70){\makebox(0.4,0.6){ $\alpha_0$}}
\put(20,70){\makebox(0.4,0.6){ $\beta_1$}}
\put(5,40){\line(1,0){10}}
\put(4,44){\line(1,1){13}}
\put(-2,45){\line(0,1){10}}
\put(2,45){\line(0,1){10}}
\put(20,50){\psline{<->}(0,-.2)(0,.2)}
}
\end{picture}} & {\small \begin{picture}(80,80)
\multiput(15,40)(20,0){2}{\circle{10}}
\put(15,20){\makebox(0.4,0.6){ $\bar{\beta_1}$}}
\put(15,10){\makebox(0.4,0.6){ ${\bf 1}$}}
\put(35,20){\makebox(0.4,0.6){ $\bar{\beta}$}}
\put(35,10){\makebox(0.4,0.6){ ${\bf 3}$}}
\put(20,40){\line(1,0){10}}
\put(50,70){\makebox(0.4,0.6){ ${\bf \N =2,d=4}$}}

\end{picture}}

\\
\hline {\small \begin{picture}(82,90)
\put(0,-20){
\multiput(0,40)(20,0){5}{\circle{10}}
\put(0,25){\makebox(0.4,0.6){ $\alpha_3$}}
\put(20,25){\makebox(0.4,0.6){$\alpha_2$}}
\put(40,25){\makebox(0.4,0.6){ $\alpha_1$}}
\put(60,25){\makebox(0.4,0.6){$\alpha_{-1}$}}
\put(77,60){\makebox(0.4,0.6){$\alpha_{-2}$}}
\put(80,25){\makebox(0.4,0.6){ $\alpha_0$}}
\put(77,80){\makebox(0.4,0.6){ $\alpha_{-3}$}}
\put(72,100){\makebox(0.4,0.6){ $\beta$}} \put(0,40){\circle*{10}}
\put(60,60){\circle{10}} \put(60,60){\circle{10}}
\put(60,100){\circle{10}}
 \put(40,40){\circle*{10}}
\put(80,40){\circle*{10}} \put(60,80){\circle{10}}
 \put(5,40){\line(1,0){10}}
\put(25,40){\line(1,0){10}}
\put(45,40){\line(1,0){10}}
\put(65,40){\line(1,0){10}}
\put(60,45){\line(0,1){10}}
\put(60,65){\line(0,1){10}}

\put(60,85){\line(0,1){10}}
}
\end{picture}} & {\small \begin{picture}(80,80)
\multiput(0,40)(20,0){5}{\circle{10}}
\put(0,20){\makebox(0.4,0.6){ $\bar{\alpha}_2$}}
\put(0,10){\makebox(0.4,0.6){ ${\bf 4}$}}
\put(35,70){\makebox(0.4,0.6){ ${\bf \N =6,d=3}$}}

\put(20,20){\makebox(0.4,0.6){$\bar{\alpha}_{-1}$}}
\put(20,10){\makebox(0.4,0.6){ ${\bf 4}$}}
\put(40,20){\makebox(0.4,0.6){$\bar{\alpha}_{-2}$}}
\put(40,10){\makebox(0.4,0.6){ ${\bf 1}$}}
\put(60,20){\makebox(0.4,0.6){$\bar{\alpha}_{-3}$}}

\put(80,10){\makebox(0.4,0.6){ ${\bf 1}$}}
\put(80,20){\makebox(0.4,0.6){$\bar{\beta}$}}
\put(60,10){\makebox(0.4,0.6){ ${\bf 1}$}} \put(5,40){\line(1,0){10}}
\put(25,42){\line(1,0){10}} \put(25,38){\line(1,0){10}}
\put(45,40){\line(1,0){10}}

\put(65,40){\line(1,0){10}}

\put(28,40){\line(1,2){5}}
\put(28,40){\line(1,-2){5}}

\end{picture}}
&{\small \begin{picture}(70,90)
\multiput(0,40)(20,0){2}{\circle{10}}
\put(0,25){\makebox(0.4,0.6){ $\alpha_1$}}
\put(20,25){\makebox(0.4,0.6){ $\alpha_2$}}
\put(10,60){\circle{10}}
\put(10,75){\makebox(0.4,0.6){ $\beta_1$}}
\put(5,40){\line(1,0){10}}
\put(3,43){\line(1,2){6}}
\put(17,43){\line(-1,2){6}}

\multiput(40,40)(20,0){2}{\circle{10}}
\put(40,25){\makebox(0.4,0.6){ $\alpha_0$}}
\put(60,25){\makebox(0.4,0.6){ $\beta_2$}}
\put(45,42){\line(1,0){10}}
\put(45,38){\line(1,0){10}}
\put(40,40){\circle*{10}}

\put(10,33){\psline{<->}(-.3,0)(.3,0)}

\end{picture}} & {\small \begin{picture}(80,80)
\multiput(15,40)(20,0){2}{\circle{10}}
\put(15,20){\makebox(0.4,0.6){ $\bar{\alpha_1}$}}
\put(15,10){\makebox(0.4,0.6){ ${\bf 2}$}}
\put(35,20){\makebox(0.4,0.6){ $\bar{\beta_1}$}}
\put(35,10){\makebox(0.4,0.6){ ${\bf 1}$}}
\put(20,38){\line(1,0){10}}
\put(20,42){\line(1,0){10}}
\put(60,40){\circle{10}}
\put(60,20){\makebox(0.4,0.6){ $\bar{\beta_2}$}}

\put(60,10){\makebox(0.4,0.6){ ${\bf 3}$}}

\put(50,70){\makebox(0.4,0.6){ ${\bf \N =2,d=3}$}}

\end{picture}}

\\

\hline {\small \begin{picture}(80,60)
\put(0,-20){
\multiput(0,40)(20,0){4}{\circle{10}}
\put(0,25){\makebox(0.4,0.6){ $\alpha_1$}}
\put(20,25){\makebox(0.4,0.6){$\alpha_2$}}
\put(40,25){\makebox(0.4,0.6){ $\alpha_3$}}
\put(60,25){\makebox(0.4,0.6){$\beta$}}
\put(0,70){\makebox(0.4,0.6){$\alpha_{5}$}}
\put(40,70){\makebox(0.4,0.6){ $\alpha_4$}}
\put(20,40){\circle*{10}} \put(40,40){\circle*{10}}
\put(40,60){\circle*{10}}
\put(0,60){\circle{10}}
\put(5,40){\line(1,0){10}}
\put(25,40){\line(1,0){10}}
\put(45,40){\line(1,0){10}}
\put(0,50){\psline{<->}(0,-.2)(0,.2)}
\put(5,60){\line(1,0){40}}
\put(40,45){\line(0,1){10}}
}
\end{picture}} & {\small \begin{picture}(80,60)
\put(0,-8){
\multiput(15,40)(20,0){2}{\circle{10}}
\put(15,20){\makebox(0.4,0.6){ $\bar{\alpha}_1$}}
\put(15,10){\makebox(0.4,0.6){ ${\bf 8}$}}
\put(35,20){\makebox(0.4,0.6){$\bar{\beta}$}}
\put(35,10){\makebox(0.4,0.6){ ${\bf 4}$}}
\put(50,60){\makebox(0.4,0.6){ ${\bf \N =5,d=4}$}}
\put(20,42){\line(1,0){10}}
\put(20,38){\line(1,0){10}}
}
\end{picture}}
&{\small \begin{picture}(70,60)
\put(10,-20){
\multiput(0,40)(0,20){2}{\circle{10}}
\put(20,40){\circle{10}}
\put(0,25){\makebox(0.4,0.6){ $\alpha_1$}}
\put(20,25){\makebox(0.4,0.6){ $\beta_1$}}
\put(0,70){\makebox(0.4,0.6){ $\beta_2$}}
\put(5,42){\line(1,0){10}}
\put(5,38){\line(1,0){10}}
\put(-2,45){\line(0,1){10}}
\put(2,45){\line(0,1){10}}
\put(6,51){\psline{<->}(.4,-.2)(0,.2)}
}
\end{picture}} & 
{\small \begin{picture}(80,60)
\put(0,-8){
\multiput(15,40)(20,0){2}{\circle{10}}
\put(15,20){\makebox(0.4,0.6){ $\bar{\alpha_1}$}}
\put(15,10){\makebox(0.4,0.6){ ${\bf 1}$}}
\put(35,20){\makebox(0.4,0.6){ $\bar{\beta_1}$}}
\put(35,10){\makebox(0.4,0.6){ ${\bf 2}$}}
\put(20,38){\line(1,0){10}}
\put(20,42){\line(1,0){10}}
\put(50,60){\makebox(0.4,0.6){ ${\bf \N =1,d=4}$}}
}
\end{picture}}

\\

\hline {\small \begin{picture}(80,75)
\put(0,-15){
\multiput(0,40)(20,0){5}{\circle{10}}
\put(0,25){\makebox(0.4,0.6){ $\alpha_1$}}
\put(20,25){\makebox(0.4,0.6){$\alpha_2$}}
\put(40,25){\makebox(0.4,0.6){ $\alpha_3$}}
\put(60,25){\makebox(0.4,0.6){$\alpha_{-1}$}}
\put(80,25){\makebox(0.4,0.6){$\beta$}}
\put(0,70){\makebox(0.4,0.6){$\alpha_{5}$}}
\put(40,70){\makebox(0.4,0.6){ $\alpha_4$}}
\put(20,40){\circle*{10}} \put(40,40){\circle*{10}}
\put(40,60){\circle*{10}}
\put(0,60){\circle{10}}
\put(5,40){\line(1,0){10}}
\put(25,40){\line(1,0){10}} \put(45,40){\line(1,0){10}}
\put(5,60){\line(1,0){40}}
\put(65,40){\line(1,0){10}}
\put(0,50){\psline{<->}(0,-.2)(0,.2)}
\put(40,45){\line(0,1){10}}
}
\end{picture}} & {\small \begin{picture}(80,75)
\put(0,-8){
\multiput(15,40)(20,0){3}{\circle{10}}
\put(15,20){\makebox(0.4,0.6){ $\bar{\alpha}_1$}}
\put(15,10){\makebox(0.4,0.6){ ${\bf 8}$}}
\put(35,20){\makebox(0.4,0.6){$\bar{\alpha}_{-1}$}}
\put(35,10){\makebox(0.4,0.6){ ${\bf 6}$}}
\put(55,20){\makebox(0.4,0.6){$\bar{\beta}$}}
\put(55,10){\makebox(0.4,0.6){ ${\bf 1}$}}
\put(50,70){\makebox(0.4,0.6){ ${\bf \N =5,d=3}$}}
\put(20,42){\line(1,0){10}}
\put(20,38){\line(1,0){10}} \put(40,40){\line(1,0){10}}
\put(23,40){\line(1,2){5}}
\put(23,40){\line(1,-2){5}}
}
\end{picture}}

&
{\small \begin{picture}(70,75)
\put(10,-15){
\multiput(0,40)(20,0){3}{\circle{10}}
\put(5,60){\makebox(0.4,0.6){ $\alpha_1$}}
\put(0,25){\makebox(0.4,0.6){ $\alpha_2$}}
\put(20,25){\makebox(0.4,0.6){$\beta$}}
\put(40,25){\makebox(0.4,0.6){ $\alpha_3$}}
\put(20,60){\circle{10}}
\put(5,40){\line(1,0){10}}
\put(25,40){\line(1,0){10}}
\put(20,45){\line(0,1){10}}
\put(20,33){\psline{<->}(-.6,0)(.6,0)}
}
\end{picture}} & {\small \begin{picture}(80,75)
\put(0,-8){
\multiput(15,35)(20,0){2}{\circle{10}}
\put(35,55){\circle{10}}
\put(20,35){\line(1,0){10}}
\put(35,40){\line(0,1){10}}
\put(15,20){\makebox(0.4,0.6){ $\bar{\alpha}_2$}}
\put(15,10){\makebox(0.4,0.6){ ${\bf 2}$}}
\put(35,20){\makebox(0.4,0.6){$\bar{\beta}$}}
\put(35,10){\makebox(0.4,0.6){ ${\bf 1}$}}
\put(48,60){\makebox(0.4,0.6){$\bar{\alpha}_1$}}
\put(48,50){\makebox(0.4,0.6){ ${\bf 1}$}}
\put(50,70){\makebox(0.4,0.6){ ${\bf \N =1,d=3}$}}
}
\end{picture}}

\\
\hline
{\small \begin{picture}(80,60)
\put(0,-15){
\multiput(0,40)(20,0){4}{\circle{10}}
\put(0,25){\makebox(0.4,0.6){ $\alpha_3$}}
\put(20,25){\makebox(0.4,0.6){$\alpha_2$}}
\put(40,25){\makebox(0.4,0.6){ $\beta$}}
\put(5,60){\makebox(0.4,0.6){ $\alpha_1$}}
\put(60,25){\makebox(0.4,0.6){ $\alpha_0$}}
\put(0,40){\circle*{10}} \put(20,40){\circle*{10}}
\put(20,60){\circle*{10}}
\put(5,40){\line(1,0){10}}
\put(25,40){\line(1,0){10}} \put(45,40){\line(1,0){10}}
\put(20,45){\line(0,1){10}}
}
\end{picture}} & {\small \begin{picture}(80,60)
\put(0,-8){
\multiput(15,40)(20,0){2}{\circle{10}}
\put(15,20){\makebox(0.4,0.6){ $\bar{\alpha}_0$}}
\put(15,10){\makebox(0.4,0.6){ ${\bf 1}$}}
\put(35,20){\makebox(0.4,0.6){$\bar{\beta}$}}
\put(35,10){\makebox(0.4,0.6){ ${\bf 4}$}}
\put(20,40){\line(1,0){10}}
\put(50,60){\makebox(0.4,0.6){ ${\bf \N =4,d=4}$}}
}
\end{picture}}
&&
 \\

\hline
{\small \begin{picture}(80,60)
\put(0,-15){
\multiput(0,40)(20,0){4}{\circle{10}}
\put(5,60){\makebox(0.4,0.6){ $\alpha_1$}}
\put(0,25){\makebox(0.4,0.6){ $\alpha_3$}}
\put(20,25){\makebox(0.4,0.6){$\alpha_2$}}
\put(40,25){\makebox(0.4,0.6){ $\alpha_{-1}$}}
\put(55,60){\makebox(0.4,0.6){ $\alpha_{0}$}}
\put(60,25){\makebox(0.4,0.6){ $\beta$}}
\put(0,40){\circle*{10}} \put(20,40){\circle*{10}}
\put(20,60){\circle*{10}} \put(40,60){\circle{10}}
\put(5,40){\line(1,0){10}}
\put(25,40){\line(1,0){10}} \put(45,40){\line(1,0){10}}
\put(20,45){\line(0,1){10}}
\put(40,45){\line(0,1){10}}
}
\end{picture}} & {\small \begin{picture}(80,60)
\put(0,-8){
\multiput(15,40)(20,0){3}{\circle{10}}
\put(15,20){\makebox(0.4,0.6){ $\bar{\alpha}_0$}}
\put(15,10){\makebox(0.4,0.6){ ${\bf 1}$}}
\put(35,20){\makebox(0.4,0.6){$\bar{\alpha}_{-1}$}}
\put(35,10){\makebox(0.4,0.6){ ${\bf 4}$}}
\put(55,20){\makebox(0.4,0.6){$\bar{\beta}$}}
\put(55,10){\makebox(0.4,0.6){ ${\bf 1}$}}
\put(50,60){\makebox(0.4,0.6){ ${\bf \N =4,d=3}$}}
\put(40,40){\line(1,0){10}}
\put(20,38){\line(1,0){10}}
\put(20,42){\line(1,0){10}}
\put(27,40){\line(-1,2){5}}
\put(27,40){\line(-1,-2){5}}
}
\end{picture}}
&

\hspace{5mm} {\small \begin{picture}(70,60)
\put(0,-15){
\multiput(0,40)(20,0){2}{\circle{10}}
\put(0,25){\makebox(0.4,0.6){ $\alpha$}}
\put(20,25){\makebox(0.4,0.6){$\beta$}}
\put(5,42){\line(1,0){10}}
\put(5,38){\line(1,0){10}}
}
\end{picture}} &

{\small \begin{picture}(80,60)
\put(0,-8){
\multiput(15,40)(20,0){2}{\circle{10}}
\put(15,20){\makebox(0.4,0.6){ $\bar{\alpha}$}}
\put(50,60){\makebox(0.4,0.6){ ${\bf \N =0,d=3}$}}
\put(15,10){\makebox(0.4,0.6){ ${\bf 1}$}}
\put(35,20){\makebox(0.4,0.6){$\bar{\beta}$}}
\put(35,10){\makebox(0.4,0.6){ ${\bf 1}$}}
\put(20,38){\line(1,0){10}}
\put(20,42){\line(1,0){10}}
}
\end{picture}}
 \\
\hline
\end{tabular}}
\end{center}
\caption{Satake diagrams of superalgebras of the magic triangle and
restricted roots.}
\end{table}

\newpage

\section*{Acknowledgments}

    We are grateful to A. Beauville, V. Kharlamov, and J. Kollar for useful
explanations and references in real algebraic geometry and to A. Keurentjes
for pointing us some errors in a preliminary version.
We would like to dedicate this work to the memory of Peter Slodowy: a 
fine mathematician and gentleman.

\end{document}